\documentstyle[aps,prd,preprint,floats,epsfig,tighten]{revtex}
\newcommand{\wfig}[4]{\begin{figure*}[htbp]\vfill\begin{center}
\mbox{\epsfig{figure=#1,width=#2}}\caption{#3}\label{#4}
\end{center}\vfill\end{figure*}}

\begin{document}
\preprint{\vbox{\hfill SLAC--PUB--7344 \\ 
                \hspace*{1cm} \hfill October 1996 \\}}
\title{
Differences between Monte Carlo models for DIS \\
at small-$x$ and the relation to BFKL dynamics
\thanks{Work supported by the Swedish Natural Science Research 
Council, contract F--PD 11264--301 
and the U.S. Department of Energy, contract DE--AC03--76SF00515.}}
\author{J.~Rathsman}
\address{Stanford Linear Accelerator Center \\
Stanford University, Stanford, California 94309, USA}

\maketitle

\begin{abstract}
The differences between two standard Monte Carlo models, 
{\sc Lepto} and {\sc Ariadne}, for deep
inelastic scattering at small-$x$ is analysed in detail. 
It is shown that the difference  arises from
a `unorthodox' suppression factor used in {\sc Ariadne}  which replaces
the normal ratio of parton densities. This gives rise to a factor that
qualitatively is similar to what one would expect from BFKL dynamics
for some observables like the energy flow and forward jets but not
for the 2+1 jet cross-section. It is also discussed how
one could use the 2+1 jet cross-section as a probe for BFKL
dynamics.
\end{abstract}

\narrowtext
%%%%%%%%%%%%%%%%%%%%%%%%%%%%%%%%%%%%%%%%%%%%%%%%%%%%%%%%%%%%%%%%%%%%%%%
Deep inelastic electron proton scattering at small-$x$ (with $x$ being
the ordinary Bjorken-$x$) has drawn much theoretical and experimental 
interest the last couple of years. Since the observation
\cite{ZEUSF2,H1F2} of the predicted \cite{derujula}
rise of the deep inelastic structure function $F_2$ at small-$x$, 
the main question has been whether the `unconventional' so called 
BFKL \cite{bfkl} dynamics can be observed or whether the interactions 
can be described by ordinary leading $\log(Q^2)$ dynamics in
perturbative QCD as given by the DGLAP \cite{dglap}
evolution equation. From the inclusive
measurement of $F_2$ it seems not 
possible to draw any such conclusions
based on the presently available data and it is also theoretically
questionable
whether it will be at all possible as long as one only considers
one observable
\cite{catani}. 

Other observables that have been proposed to see effects
of BFKL dynamics are the transverse energy flow \cite{eflow-martin} and
forward jets \cite{mueller}. One problem in looking for possible
effects of BFKL dynamics is that so far there has only been analytical
calculations on parton level and the hadronisation effects are
difficult to estimate. In this respect the forward jets analysis is
more promising since the hadronisation corrections are expected to be 
small for a jet cross-section. However, the forward jets are close
to the target remnant jet which makes it difficult to disentangle the 
two.

So far there exists no Monte Carlo (MC) model based on the BFKL equation
even though there has recently been significant progress in creating
such a model \cite{ldc} based on the CCFM equations \cite{ccfm} 
which interpolate smoothly between the BFKL and DGLAP equations.
In the mean time the {\sc Ariadne} \cite{ariadne} MC which
is based on the colour dipole model for DIS \cite{cdmdis} has
often been used to estimate the possible effects of BFKL dynamics.
The argument for this has been that it also contains parton emissions 
which are non-ordered in transverse momentum. However,
the present paper will show that the difference between {\sc Ariadne}
and ordinary DGLAP evolution 
is mainly due to a `unorthodox' suppression factor used
for additional emissions from dipoles connected with the proton remnant.
Comparing the results from the {\sc Ariadne} MC with data has also
shown rather good agreement for the transverse 
energy flows and forward jets
whereas a pure DGLAP Monte Carlo fails. There are, however, large 
hadronisation uncertainties and it has been shown that the 
{\sc Lepto} \cite{lepto} MC model 
based on leading order
(LO) QCD matrix-elements and leading logarithmic 
corrections in the form of parton showers according to 
the DGLAP evolution equation supplemented 
with non-perturbative hadronisation effects
can also describe the data on transverse energy flow 
\cite{EIR}.

An observable which at first sight might seem more or less uncorrelated
to the question of whether one observes BFKL dynamics or not is the
2+1 jet cross-section. Since this is a genuinely hard 
process it should be well described by conventional perturbative
QCD as long as the more complicated forward region is excluded. 
However this is not the case. 
As argued in this paper the 2+1 jet
cross-section should exhibit features of BFKL dynamics when the
mass $\hat{s}$ of the jet-system becomes much larger than
the momentum transfer $\hat{t}$ from the incoming parton.
In other words one should be able
to observe effects of BFKL dynamics as the rapidity
difference ($\Delta{y}=\ln(\hat{s}/|\hat{t}|)$) 
between the two jets increases. 
This is just the same kind of mechanism as in dijet production
in hadron collisions as proposed by Mueller and Navelet
\cite{mueller-navelet}.
There has also been a proposal \cite{askew} to
use the 2+1 jet cross-section for small jet-systems as a probe of
BFKL dynamics since this would essentially probe the $x$-distribution
of gluons at small $x$ where effects of BFKL dynamics 
should be possible to observe. A complication with this proposal 
is that the gluon density is not an observable and the constraints 
from data on what it should be in a DGLAP evolution scenario are small
especially if one takes into the account the theoretical uncertainties
in defining the gluon density. There are also experimental 
difficulties in finding jets for small mass jet-systems.

At large values of photon virtuallity $Q^2$ 
(typically $Q^2>100$ GeV$^2$) the measured jet cross-section has 
been shown to be in good agreement 
\cite{jetzeus,jeth1,alphash1,alphaszeus} with the 
Monte Carlo 
model {\sc Lepto} and next-to-leading order
(NLO) perturbative QCD calculations using programs such
as {\sc Disjet} \cite{disjet} and 
{\sc Projet} \cite{projet}. In addition, the {\sc Ariadne} MC also
describes the $y_{cut}$ dependence (where $y_{cut}$ is the cut 
used in the JADE jet-definition, $m_{ij}^2 \leq y_{cut}W^2$) of 
the data with a similar accuracy to the one found for
{\sc Lepto} whereas the $Q^2$-dependence is not so well described
\cite{jeth1,alphash1}.
These cross-section measurements for large $Q^2$
have also been used for the 
extraction of $\alpha_s$ and the observation of its running
\cite{alphash1,alphaszeus}.

However, if one instead looks at the small-$x$ and small $Q^2$ 
region, 
where $x=Q^2/2P \cdot q$ with $q$ and $P$ being the photon and proton
four momenta respectively, but still
avoiding the more complicated forward region, the picture does not
seem to be so clear anymore. There is still fairly good agreement 
between {\sc Lepto} and data \cite{jeth1,alphash1} but
the two MC models give quite different results.
As an example Table \ref{tab:me} gives 
the 2+1 jet cross-section on matrix-element level 
as calculated with {\sc Lepto} 6.5, {\sc Ariadne} 4.08 and 
{\sc Mepjet} 1.1 
\cite{mepjet} which is a NLO perturbative QCD calculation.
The kinematical region used is: $0.0001<x<0.001$, 
$0.05<y_B<0.7$ ($y_B=q \cdot P /k \cdot P$ with k being the 
momentum of the incoming lepton),
$5.0 < Q^2 <70.0 $ GeV$^2$ and the MRS A
parton distributions \cite{mrsa} have been used. 
The jets have been defined
using the cone-algorithm in the hadronic center of mass 
system ({\it hcms})
with a cone-size of $R=\sqrt{\Delta \eta^2 + \Delta \phi^2}=1$ and
requiring the jets to have transverse momenta
$p_\perp>5$ GeV and $0.1 <z< 0.9$
(where $z=P \cdot j_i/P \cdot q$ with $j_i$ being the jet-momenta)
to avoid the forward region. This gives jets which have 
pseudo-rapidities in the photon-hemisphere in the {\it hcms} and
$-2<\eta<2$ in lab, thus avoiding the forward region. 
The cut in $z$ also decreases the effects from parton showers
on the 2+1 jet cross-section to less than 10 percent \cite{alphash1}.

\begin{table}[htb]
\caption{Jet cross-sections for $0.0001<x<0.001$, $0.05<y_B<0.7$
and $5.0 < Q^2 <70.0 $ GeV$^2$. 
\label{tab:me}}
\begin{tabular}{ddddd}
 Program         & $\mu_F^2$, $\mu_R^2$          & $\sigma$ [$nb$]\\
\hline    
 {\sc Ariadne}      & $Q^2$, ${p}_\perp^2$          & 6.38    \\
 {\sc Lepto}        & $Q^2$                         & 1.52    \\
 {\sc Mepjet}  LO   & $Q^2$                         & 1.49    \\
 {\sc Mepjet}  LO   & $Q^2$,$(\Sigma{p}_\perp/2)^2$ & 1.19    \\
 {\sc Mepjet}  LO   & $(\Sigma{p}_\perp/2)^2$ & 1.45    \\
 {\sc Mepjet}  NLO  & $Q^2$                   & 0.89$\pm$0.06 \\
 {\sc Mepjet}  NLO  & $(\Sigma{p}_\perp/2)^2$ & 0.95$\pm$0.04 
\end{tabular}
\end{table}

The first thing to note is that the 2+1 jet cross-section predicted by
the {\sc Ariadne} MC is about four times larger than the one from 
{\sc Lepto}. Comparing with the LO result from {\sc Mepjet} 
using $Q^2$ as renormalisation and factorisation
scale shows good agreement with {\sc Lepto}
whereas the LO result from {\sc Mepjet} with the same factorisation 
and renormalisation scales as in {\sc Ariadne} differs a factor 
five from the {\sc Ariadne} result. 
Clearly the LO 2+1 jet 
cross-section in {\sc Ariadne} does not agree with the result 
from LO perturbative QCD. 
The second thing to note is that the NLO corrections are quite 
large and that they in fact are negative such that the NLO 
cross-section is smaller than the LO one. Of course the question of
choosing an appropriate renormalisation scale and factorisation
scheme and scale has to be addressed if one wants to compare the NLO
cross-sections with data but for the present purposes, i.e. to see
that the NLO-corrections cannot increase the cross-section by a large
factor it is not necessary to pursue these theoretical 
uncertainties further (for a discussion of the renormalisation scale 
uncertainty see for example \cite{scale}).

\wfig{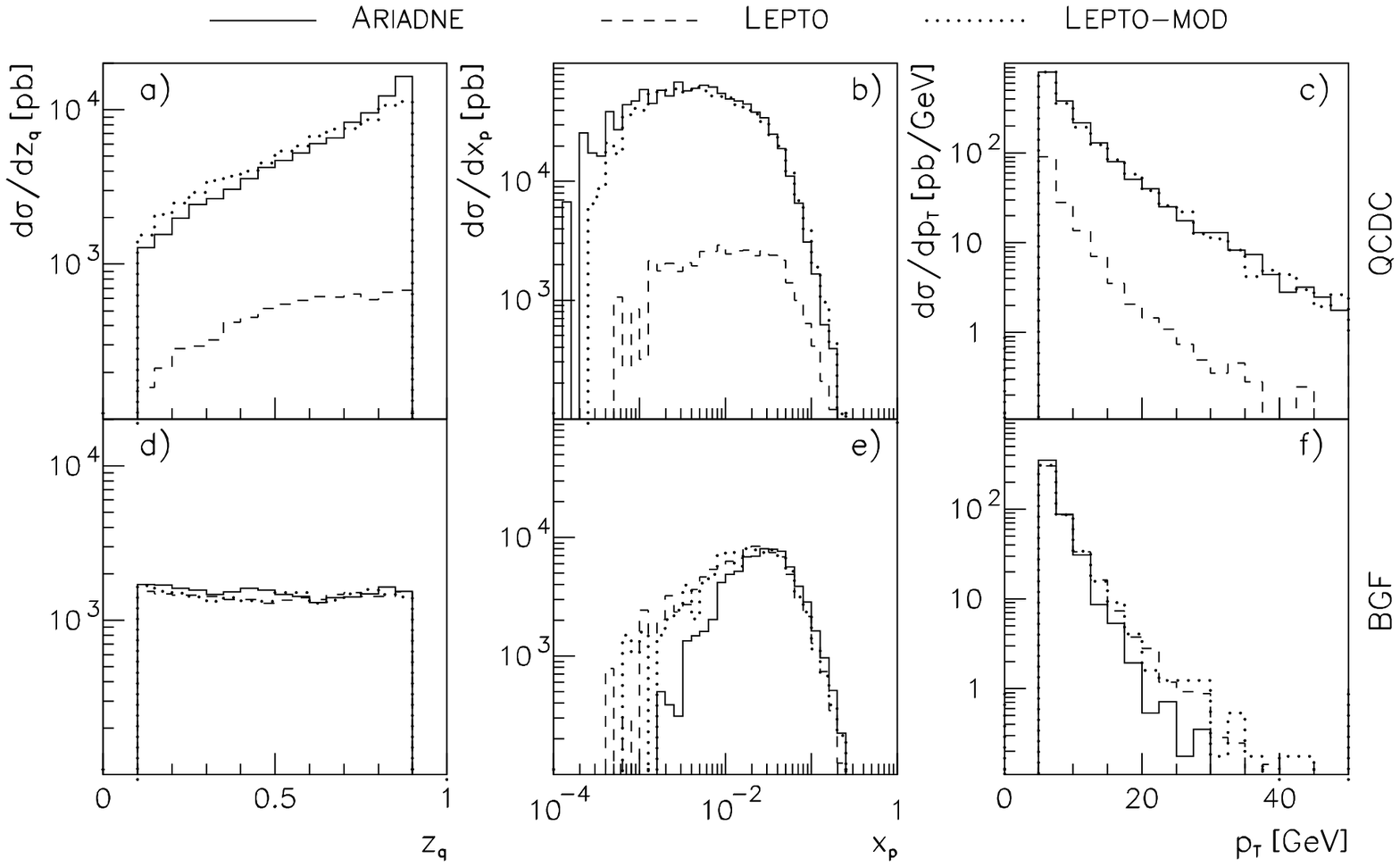}{17cm}{The $z_q$, $x_p$ and $p_\perp$ distributions
from the LO matrix-elements in {\sc Ariadne} 4.08 (full line), 
{\sc Lepto} 6.5 (dashed line) and the modified version of 
{\sc Lepto} (dotted line) for the two basic processes 
QCD-compton (QCDC) and Boson Gluon Fusion 
(BGF).}{fig:me}

To see in more detail what the difference between {\sc Ariadne} 
and LO perturbative QCD is (as represented by {\sc Lepto}), it
is instructive to look at the $z$, $x_p$ and $p_\perp$ distributions
which are given in Fig.~\ref{fig:me}. Here $x_p=x/\xi$ where $\xi$
is the longitudinal momentum share of the parton entering the hard 
interaction. For clarity the cross-section
has been divided into two different parts, QCD-compton (QCDC) and
Boson Gluon Fusion (BGF), depending on the underlying hard process
as illustrated in Fig.~\ref{fig:2plus1}.
As can be seen from Fig.~\ref{fig:me}, the BGF part agrees well 
between 
the two programs taking into account the different renormalisation 
scales, but for the QCDC part the difference is dramatic and requires 
some other explanation.

\wfig{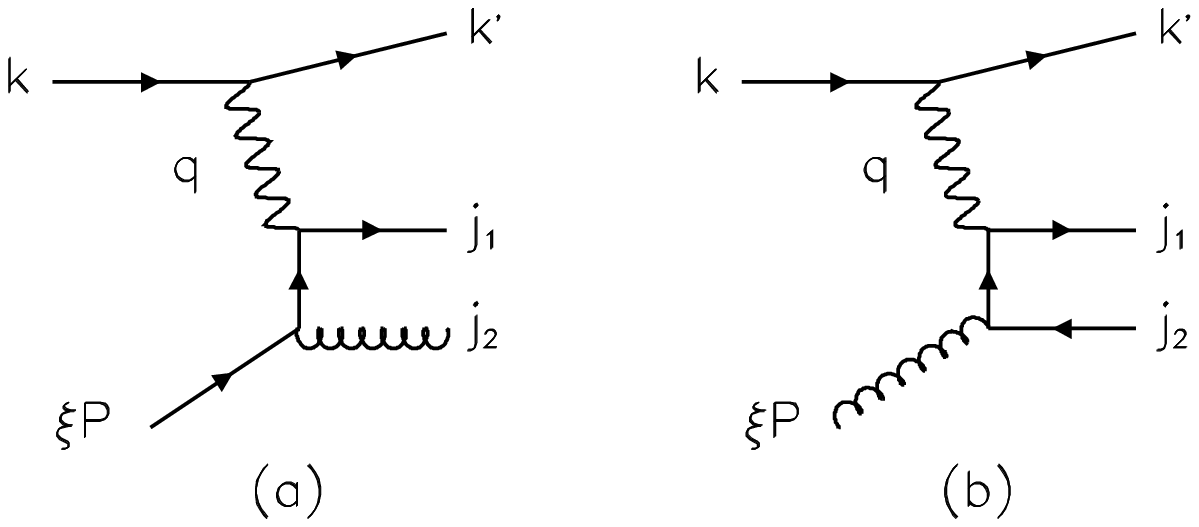}{10cm}{Illustration of the two leading order 2+1 jet
processes, QCD-compton (a) and Boson Gluon Fusion (b).}{fig:2plus1}

In fact, this difference should come as no surprise. The colour dipole
model for DIS does not use the pure LO perturbative QCD 
matrix-element for the QCDC process but rather an approximation.
For a given $x$ and $Q^2$ the cross-section can (assuming single photon 
exchange) be written as,
\begin{equation}
\frac{d\sigma_{QCDC}(x,Q^2,z,x_p)}{\sigma_{tot}(x,Q^2)}=
\frac{C_F\alpha_s}{2\pi}\frac{f_q(\xi,Q^2)}{f_q(x,Q^2)}
g(x,Q^2,z,x_p)\frac{dx_p}{x_p} dz
\end{equation}
where $f_q$ is the sum of the quark densities in the proton and 
$g(x,Q^2,z,x_p)$ is a simple function which is not needed for the
present discussion.
In the colour dipole model for DIS the ratio of the parton densities
is replaced by a suppression factor due to the extended 
proton remnant,
\begin{equation}
\frac{f_q(\xi,Q^2)}{f_q(x,Q^2)} \rightarrow 
\Theta \left(\frac{W}{e^y+(p_\perp/\mu)^a e^{-y}}-p_\perp\right).
\end{equation}
where $y$ is the rapidity of the emitted parton in the {\it hcms}
and $W$ is the mass of the hadronic system ($W^2=(P+q)^2$). 
This limits the transverse momentum $p_\perp$ of the emitted gluon
which corresponds to that only a fraction $({\mu}/{p_\perp})^a$
of the proton remnant takes part in the emission. Normally one
uses $a=1$ but other choices are also possible.
Now, for small-$x$ the suppression factor used in the 
colour dipole model for DIS
starts to differ substantially from the ratio of the parton densities.
In the region of interest $x$ is very small and $\xi$ is moderately 
small
(typically $10^{-2}-10^{-1}$) so that this ratio goes essentially as 
$x_p^{1+\lambda}$ if one assumes that
$f_q(x)\propto x^{-1-\lambda}$. For recent parton distributions 
$\lambda$ is in the order of 0.2 to 0.3 so this gives a strong 
suppression of small $x_p$ in the LO perturbative QCD formula.

To see that it really is this suppression factor that gives the
difference a "toy-model" version of {\sc Lepto} has been constructed.
In this modified version of {\sc Lepto} the QCDC matrix-element
has been multiplied by a phenomenological factor 
$x_p^{-b}$ where $b=d-0.05\ln{1/x_p}$ with $d$ being a function of $x$.
The value of $d$ was obtained by a fit to the ratio of
$d\sigma_{QCDC}/dx_p$ from {\sc Ariadne} and {\sc Lepto}
for fixed $x$ and $Q^2$. The resulting powers turn out to
be easily parameterised
by $d=0.2(1-\log_{10}x)$ in the region of 
interest ($10^{-4}<x<10^{-3}$). 
As can be seen in Fig.~\ref{fig:me} this "toy-model" 
reproduces the distributions from {\sc Ariadne} to a very good 
approximation, especially the $p_\perp$-distribution. 
Thus, the difference between the QCDC cross-section in
{\sc Lepto} and {\sc Ariadne} can be explained by a factor $x_p^{-b}$ 
which results from using the suppression factor 
$\Theta \left(\frac{W}{e^y+(p_\perp/\mu)^a e^{-y}}-p_\perp\right)$
instead of $f_q(\xi,Q^2)/f_q(x,Q^2)$.
As an aside, it is interesting to note that for the BGF-part of 
the jet cross-section, the suppression factor used in {\sc Ariadne}
is indeed the corresponding ratio of the parton densities.

The important thing to note is that all emissions in {\sc Ariadne}, 
where the proton remnant is part of the dipole, are
treated in the same way as the first emission in the 
QCDC matrix-element. To be more precise the probability for
an extra emission, when the proton remnant is part of the dipole, is 
given by \cite{leif}
\begin{equation}
dP\propto\frac{4\alpha_s}{3\pi}
\Theta \left(\frac{W}{e^y+(p_\perp/\mu)^a e^{-y}}-p_\perp\right)
\frac{dp_\perp^2}{p_\perp^2}dy.
\end{equation}
This means that the ratio of 
parton densities ${f_a(x_a,Q^2)}/{f_b(x_b,Q^2)}$ does not enter 
as they would in a traditional backwards evolution initial 
state parton shower where,
\begin{equation}
dP_{a\to bc}\propto\frac{4\alpha_s}{3\pi}\frac{f_a(x_a,Q^2)}{f_b(x_b,Q^2)}
\frac{dp_\perp^2}{p_\perp^2}dy.
\end{equation}
Instead the boundary condition
present from the proton is taken into account by the extra cut-off 
in transverse momenta for emitted gluons due to the extendedness 
of the proton remnant. A direct comparison
with a traditional initial state parton shower 
is complicated by the fact that
the emissions in the colour dipole model for DIS
are not ordered in $x$.
But even so, it is evident that the difference in the suppression 
factor gives a large effect for the rest of the emissions just as 
for the QCDC matrix-element. There are two
main effects. First of all the probability for an emission is 
increased
and secondly the emissions become harder, in the sense that they 
get higher $p_\perp$, just in the same way as the QCDC part
of the 2+1 jet cross-section increases and the $p_\perp$ spectrum 
becomes harder (see Fig.~\ref{fig:me}c). 

To estimate the magnitude of the effects on the transverse energy flow
due to the `unorthodox' suppression factor
the modified version of {\sc Lepto} with
the increased QCDC cross-section has been used to calculate
the transverse energy flow. The results are shown in 
Fig.~\ref{fig:eflow} where data from H1 \cite{H1eflow} are compared
with {\sc Ariadne}, {\sc Lepto} and the modified version of {\sc Lepto}.
One should here note that 
to be able to compare with more or less pure DGLAP dynamics the
soft colour interactions and the new seaquark treatment in {\sc Lepto} 
have been shut off for both {\sc Lepto} versions 
(see \cite{EIR} for more details).

As can be seen from Fig.~\ref{fig:eflow} the modified version of 
{\sc Lepto} interpolates nicely between the {\sc Ariadne} result in
the photon hemisphere where the first emission is important and 
the DGLAP-{\sc Lepto} result in the proton hemisphere where higher
order emissions are important. 
In other words, the modified matrix-element increases the transverse
energy flow to the same level as in Ariadne in the photon hemisphere,
whereas in the proton hemisphere there is no difference between the
two versions of {\sc Lepto} since the same initial state parton shower
is used. It would also be possible to modify 
the initial state parton shower in 
{\sc Lepto} in the same way as the matrix-element 
(essentially by multiplying the splitting function 
$P_{q \to qG}(z)$ with a factor $z^{-b}$) which
would lead to an increased transverse energy flow in the
proton hemisphere. However, already from the result based on the
modified QCDC matrix-element one can conclude that the 
difference in transverse energy flow between DGLAP-{\sc Lepto} and
{\sc Ariadne} is mainly due to the `unorthodox' suppression factor 
used in {\sc Ariadne} instead of the ratio of parton densities and
not due to the difference in $p_\perp$-ordering.

The following question then arises. Has this difference in
suppression factor
anything to do with BFKL dynamics? As it turns out, 
the answer depends on which
observable one is interested in. Starting with the 2+1 jet 
cross-section one first notes that the relevant scales for the onset
of BFKL dynamics are the momentum transfer $\hat{t}$ from the incoming 
gluon and the mass $\hat{s}$ of the jet-system. One expects effects
of BFKL dynamics when the rapidity
difference $\Delta{y}=\ln(\hat{s}/|\hat{t}|)$ becomes large so that
$\alpha_s \Delta{y} \sim {\cal O}(1)$ which indicates that
resummation is necessary.
In the {\it hcms} the relevant ratio becomes 
$|\hat{t}|/\hat{s} \simeq p_\perp^2/\hat{s} = z(1-z)$.
This is quite different from the increase with decreasing
$x_p=Q^2/(Q^2+\hat{s})$ given by {\sc Ariadne}. One also notes that 
typical $\xi$ values are of the order of $10^{-2}-10^{-1}$ which is
not very small and therefore one
does not expect any effects from the BFKL dynamics in the parton
densities.

\wfig{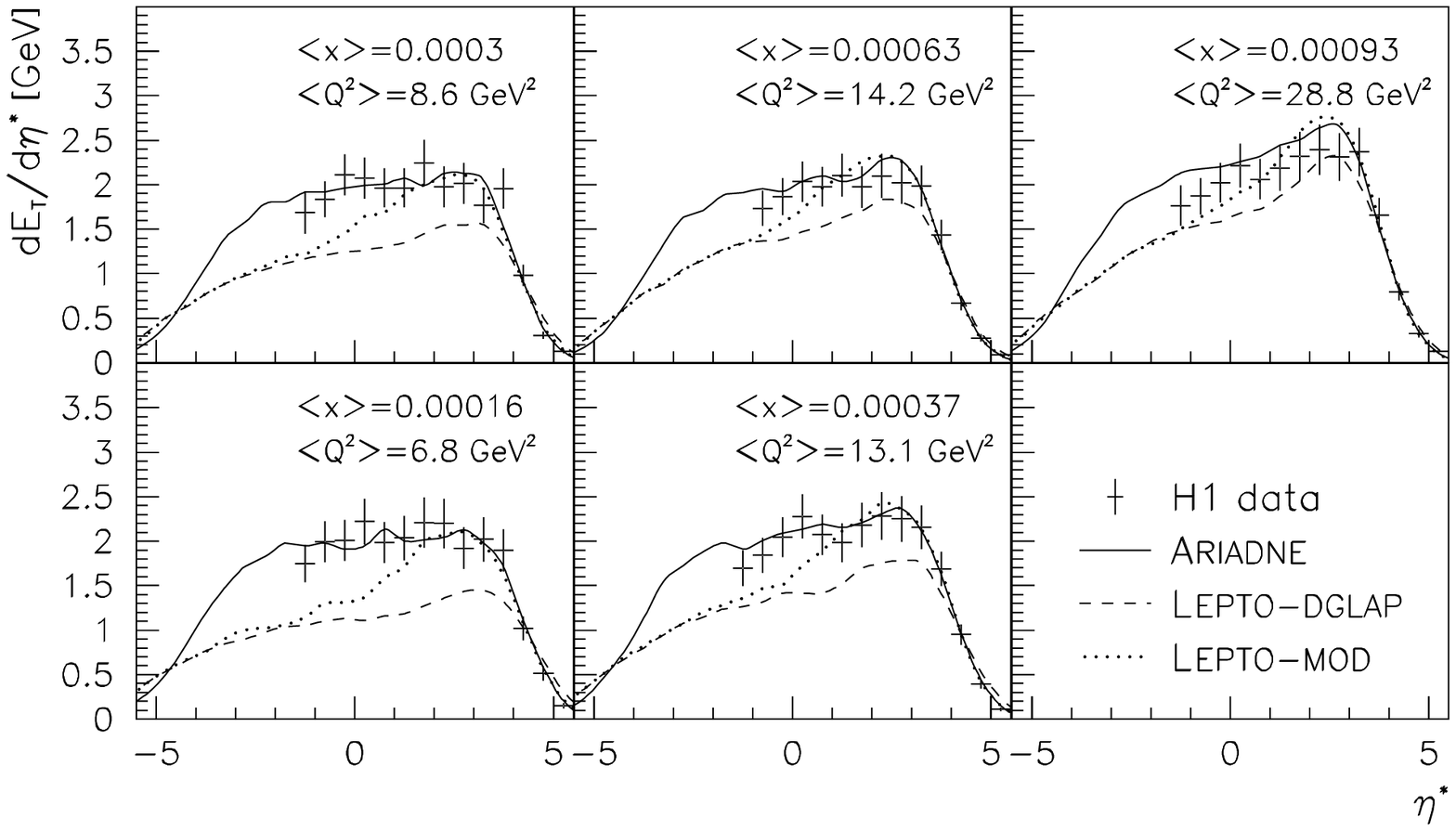}{17cm}{The transverse energy flows for different 
$x$ and $Q^2$ values in the hadronic cms comparing {\sc Ariadne}
(full line), {\sc Lepto} with pure DGLAP (dashed line) and the 
modified version of {\sc Lepto} (dotted line) with data from H1 
\protect\cite{H1eflow}. (The plot has been generated using the 
HzTool package \protect\cite{hztool}).}{fig:eflow}
\wfig{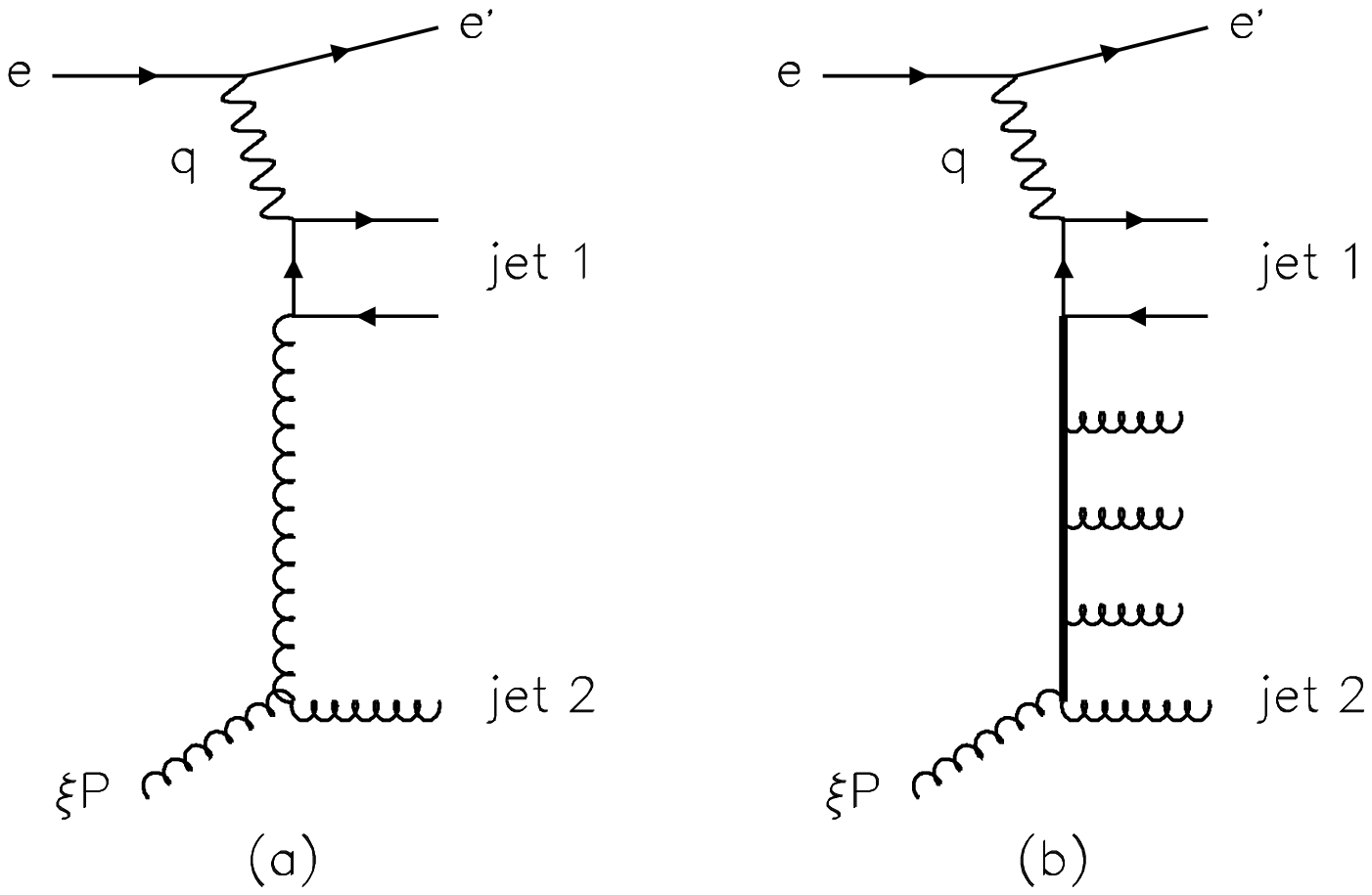}{11cm}{Diagrams with enhancement from BFKL dynamics 
in the large $\Delta{y}$ limit: (a) lowest order diagram
(b) all orders diagram with the reggeized gluon indicated with a 
thick line. In addition there is a similar diagram where
the incoming and outgoing gluon is replaced by a quark.}{fig:bfkl}

The dominating diagram for 2+1 jet production with enhancement
from the resummation of soft gluons according to the BFKL
prescription is depicted in Fig.~\ref{fig:bfkl} where the two 
partons from
the quark box form one jet. In addition to this process there is also
a similar one where the incoming and outgoing gluon is replaced by a 
quark.
At sufficiently large rapidity difference $\Delta{y}$
the contribution from this kind of diagrams should be dominating
thanks to the t-channel gluon exchange. This is the same kind
of diagram that has been analysed for the forward jet production
\cite{mueller} but with the important difference that two jets
are required. So instead of having $k_\perp$ of the forward jet
being of the same order as $Q$ for the BFKL dynamics to be valid,
two jets with the same $p_\perp$ are required. In a sense this is 
closer to the original proposal by 
Mueller and Navelet \cite{mueller-navelet} to look for BFKL
effects in dijet-production in hadron collisions.

In a BFKL calculation of the 2+1 jet cross-section one would take the
large $\Delta{y}=\ln(\hat{s}/|\hat{t}|)$ limit of the
2+1 jet cross-section and reggeize the t-channel gluon
which would give a cross-section of the type \cite{vittorio}
\begin{equation}
\frac{d\sigma}{d\xi d\Delta{y}} \propto K_{BFKL}
\left.\frac{d\hat{\sigma}}{d\xi d\Delta{y}}
\right|_{large \Delta{y}}
\left[\xi f_g(\xi,\mu^2)+\frac{C_F}{N_C}\xi f_q(\xi,\mu^2)\right],
\end{equation}
where the effective parton density \cite{combridge-maxwell}, 
$f_g(\xi,\mu^2)+\frac{C_F}{N_C}f_q(\xi,\mu^2)$, 
reflects the domination of t-channel gluon exchange
for large $\Delta{y}$. 
The factorisation scale $\mu$ is arbitrary but should be of the
order of transverse momentum cut-off for the jets, $p_{\perp,\min}$.

In a leading order calculation the enhancement factor is equal to one
so by forming the ratio between 
the BFKL calculation and the fixed order calculation for fixed $\xi$
one isolates the  $K_{BFKL}$-factor.
Thus, as a probe for BFKL dynamics,
it should be possible to use  the 2+1
jet cross-section as a function of the rapidity difference $\Delta{y}$
between the jets in the {\it hcms} (for fixed
$\xi$) and see whether
the data start to deviate exponentially from the NLO fixed order 
perturbative QCD predictions
when $\Delta{y}$ becomes large. To be able to
do a quantitative analysis one would of course need to actually
calculate the proposed cross-section. 

Integrating over $\hat{t}$ from $-p_{\perp,\min}^2$ 
the asymptotic form of the enhancement factor will be \cite{vittorio}
\begin{equation}
K_{BFKL}=\frac{\exp(\alpha_sC_A\Delta{y}4\ln{2}/\pi)}
{\sqrt{7\alpha_sC_A\zeta_3\Delta{y}/2}}.
\end{equation}
Using the standard value for $\alpha_sC_A4\ln{2}/\pi=0.5$ this
gives,
\begin{equation}
K_{BFKL} \propto \exp(\alpha_sC_A\Delta{y}4\ln{2}/\pi) =
\left(\frac{\hat{s}}{p_{\perp,\min}^2}\right)^{0.5}. 
\end{equation}
For forward jets $-\hat{t}\simeq Q^2$ is required and thus 
$\Delta{y} \simeq \ln(1/x_p)$ such that
$K_{BFKL} \propto\left(\frac{1}{x_p}\right)^{0.5}$ which also
is consistent with the results from \cite{fjet-bartels}.

Returning to the 
cross-section for the QCDC process used in 
{\sc Ariadne}, it can be written in the following way,
\begin{eqnarray}
\frac{d\sigma_{QCDC}(x,Q^2,z,x_p)}{\sigma_{tot}(x,Q^2)} & = &
\frac{C_F\alpha_s}{2\pi}
\frac{f_q(\xi,Q^2)}{f_q(x,Q^2)}
\frac{\Theta \left(\frac{W}{e^y+(p_\perp/\mu)^a e^{-y}}-p_\perp\right)}
{\frac{f_q(\xi,Q^2)}{f_q(x,Q^2)}}
g(x,Q^2,z,x_p)
\frac{dx_p}{x_p} dz \nonumber \\
 & \simeq & \frac{C_F\alpha_s}{2\pi}\frac{f_q(\xi,Q^2)}{f_q(x,Q^2)}
\left(\frac{1}{x_p}\right)^b g(x,Q^2,z,x_p) 
\frac{dx_p}{x_p} dz
\end{eqnarray}
with $b$ being in the order of $0.5-1.0$ for small $x$.
Qualitatively this gives a $K$-factor $x_p^{-b}$ which is 
similar to the one in the forward jet BFKL calculation
\cite{fjet-bartels}.
This also explains why {\sc Ariadne}
agrees quite well with data on forward jets \cite{H1eflow}
just in the same way as the forward jet BFKL calculation 
\cite{fjet-bartels} agrees quite well with data.

Even though there is some qualitative agreement between the 
colour dipole model
and a BFKL calculation for forward jets there are
quantitative differences which are important:
{\it (i)} the colour dipole model has an enhancement already 
for the first emission whereas the BFKL enhancement only 
comes in at higher order as depicted in Fig.~\ref{fig:bfkl},
{\it (ii)} the factor giving the enhancement from the BFKL dynamics
multiplies the cross-section both for incoming gluon and incoming
quark,
{\it (iii)} the BFKL enhancement is in the hard cross-section whereas
in the colour dipole model the enhancement comes from replacing
the ratio of parton densities with a cut-off in transverse momentum
for the emitted gluon.
It should also be stressed that
for the 2+1 jet cross-section the {\sc Ariadne} prediction gives an 
even larger enhancement of the 
cross-section than one would expect from BFKL dynamics.

In summary this paper has shown the underlying dynamical reason for 
the differences in 2+1 jet cross-sections, forward jet 
cross-sections  and transverse energy flow
between the DGLAP based {\sc Lepto} Monte Carlo
and the {\sc Ariadne} Monte Carlo based on the colour dipole model
for DIS. The dominating difference is the `unorthodox' suppression
factor used instead of the ratio of parton densities
and not the $p_\perp$-ordering.
The difference can essentially be parameterised by a factor
$x_p^{-b}$, where $b$ is in the order of $0.5-1.0$ for small $x$,
which multiplies the cross-section for additional emissions in 
{\sc Ariadne}.
For the forward jet production and the transverse energy flow 
this factor $x_p^{-b}$ resembles qualitatively what one gets 
from a BFKL calculation where $b=\alpha_sC_A4\ln{2}/\pi\simeq0.5$.
However for the 2+1 jet production one rather expects an enhancement
from BFKL dynamics as a function of the rapidity 
difference $\Delta{y}=\ln(\hat{s}/|\hat{t}|)$ between the jets.
Thus, it should be possible to use the 2+1 jet cross-section 
for fixed $\xi$ and its dependence on the
rapidity difference as a probe for BFKL dynamics.

%%%%%%%%%%%%%%%%%%%%%%%%%%%%%%%%%%%%%%%%%%%%%%%%%%%%%%%%%%%%%%%%%%%%%%%
\acknowledgements
I would like to thank Anders Edin, Gunnar Ingelman and Leif L\"onnblad
for useful communications and J\"urgen Spiekerman and G\"unther
Grindhammer for bringing the discrepancy between {\sc Lepto} and {\sc
Ariadne} for the 2+1 jet cross-section to my attention which started
this investigation.

%%%%%%%%%%%%%%%%%%%%%%%%%%%%%%%% REFERENCES %%%%%%%%%%%%%%%%%%%%%%%%%%%

\end{document}